\documentclass[twocolumn,preprintnumbers,amsmath,amssymb,floatfix,superscriptaddress,nofootinbib]{revtex4}

\usepackage{amsmath,hyperref,amssymb}
\usepackage{graphicx}
\usepackage{dcolumn}
\usepackage{bm}
\usepackage{verbatim}
\usepackage{tabularx}
\usepackage{slashed}
\usepackage{cancel}
\usepackage[boxsize=2em,aligntableaux=center]{ytableau}
\usepackage{float}

\def\be{\begin{equation}}
\def\ee{\end{equation}}
\def\bea{\begin{eqnarray}}
\def\eea{\end{eqnarray}}

\begin{document}

\vspace*{-30mm}

\title{Scalars Gliding Through an Expanding Universe}

\author{Anson Hook}
\email{hook@umd.edu}
\affiliation{Maryland Center for Fundamental Physics, Department of Physics, University of Maryland, College Park, MD 20742, U.S.A.}

\author{Gustavo Marques-Tavares}
\email{gusmt@umd.edu}
\affiliation{Maryland Center for Fundamental Physics, Department of Physics, University of Maryland, College Park, MD 20742, U.S.A.}
\affiliation{Department  of  Physics  and  Astronomy,  Johns  Hopkins  University,  Baltimore,  MD  21218, U.S.A.}

\author{Yuhsin Tsai}
\email{yhtsai@umd.edu}
\affiliation{Maryland Center for Fundamental Physics, Department of Physics, University of Maryland, College Park, MD 20742, U.S.A.}

\vspace*{1cm}

\begin{abstract} 

In this article we investigate the effects of single derivative mixing in massive bosonic fields. In the regime of large mixing, we show that this leads to striking changes of the field dynamics, delaying the onset of classical oscillations and decreasing, or even eliminating, the friction due to Hubble expansion. We highlight this phenomenon with a few examples.  In the first example, we show how an axion like particle can have its number abundance parametrically enhanced.  In the second example, we demonstrate that the QCD axion can have its number abundance enhanced allowing for misalignment driven axion dark matter all the way down to $f_a$ of order astrophysical bounds.  In the third example, we show that the delayed oscillation of the scalar field can also sustain a period of inflation. In the last example, we present a situation where an oscillating scalar field is completely frictionless and does not dilute away in time.

\end{abstract}

\maketitle

\section{Introduction}

Light scalar fields are present in many scenarios of physics beyond the Standard Model. In the early Universe scalar fields are  generically displaced from the minimum of their potential, and therefore lead to a new form of energy that evolves dynamically during the history of the Universe. They might play many different roles in cosmology, such as dark matter~\cite{Abbott:1982af,Dine:1982ah,Preskill:1982cy}, dark energy~\cite{Ratra:1987rm},  moduli fields determining the value of Standard Model parameters~\cite{Damour:1994zq,Arvanitaki:2016xds} and as dynamical solutions to the Hierarchy or cosmological constant problems~\cite{Graham:2015cka,Graham:2019bfu}.

The dynamics of scalar fields in the early Universe is largely determined by the form of their potential and by the Hubble expansion. In this article we study how their evolution can be dramatically changed by introducing mixing with another massive field via a single derivative interaction~(see Refs.~\cite{Maleknejad:2011jw,Adshead:2012kp,Dimastrogiovanni:2012st,Adshead:2016iix} for other work studying single derivative interactions). As we will show, such a mixing leads to two striking features: it delays the beginning of the oscillations of the scalar field; and it also changes the dilution of its energy density due to the expansion of the Universe.

We focus most of our discussion in examples where a scalar mixes with a massive spin one field. As a case study we present a scenario where the QCD axion and a massive dark photon mix in a background magnetic field. This can lead to a significant enhancement of the final axion density, enlarging the parameter space for which axions make up all of the observed dark matter density. We also briefly discuss an example with only scalar fields, in which the energy density in the scalar field remains constant even after the field starts oscillating.

\section{A simple example}

We are interested in mixing scalars with other (bosonic) degrees of freedom via mixing terms that involve a single derivative. This requires the presence of a background field that breaks Lorentz invariance. To concentrate on the basic idea we will first study an example with a scalar $\phi$, a massive dark photon $A_{d,\mu}$, a massless photon $A_\mu$, and ignore the effects of gravity.  The Lagrangian we consider is (see also \cite{Kaneta:2016wvf,Kaneta:2017wfh,Choi:2018dqr}, where the same Lagrangian was considered in a different regime)
\bea
\mathcal{L} =  \mathcal{L}_\text{kinetic} - \frac{1}{2} m^2 \phi^2 + \frac{1}{2} M^2 A_d^\mu A_{d,\mu} + \frac{\phi}{2 f} F_d \tilde F .
\label{eq:lagrangian-axion}
\eea
We take $A$ to have a time-independent, uniform magnetic field $B$ in the $\hat x$ direction.  As we take the $B$-field to be constant in space we can focus on solutions that are also constant in space, and can ignore spatial derivatives (we discuss the changes to this scenario due to an inhomogeneous magnetic field in the supplementary material~\cite{supplement}).~\footnote{
	We cannot achieve a similar scenario with a coupling only to dark photons (i.e. $\phi F_d \tilde F_d$). The mechanism requires mixing with a massive vector which also leads to the dark magnetic field acquiring time variations on scales of order $M^{-1}$ and spatial variations on the scales of order $E_D/(B_D M)$.} When the solutions are spatially homogeneous, it can be easily shown that the scalar $\phi$ and the massive vector $A_d$ do not source the massless vector field, i.e $\delta A=0$ (where $\delta A$ are the fluctuations of $A$ on top of the $B$-field) at all times and its dynamics can be ignored.  The remaining equations of motion in this background are
\bea
\label{Eq:simple}
	\ddot \phi + \Gamma_\phi \dot \phi &=& - m^2 \phi + \frac{B}{f} \dot A_{d,x} \\
	\ddot A_{d,x} + \Gamma_{A_d}  \dot A_{d,x} &=& - M^2 A_{d,x} - \frac{B}{f} \dot \phi 
\label{Eq:simple2}
\eea
where we have included $\Gamma_\phi$ and $\Gamma_{A_d}$ to represent generic friction terms.  

Focusing on the small friction regime, in which the system is underdamped, we first solve Eqs.~(\ref{Eq:simple},~\ref{Eq:simple2}) exactly (ignoring friction) and find the oscillation frequencies
\bea
\label{eq:omega_s}
\omega_s^2 &=& \frac{\alpha - \sqrt{\alpha^2 - 4 m^2 M^2}}{2} \approx \frac{m^2 M^2}{B^2/f^2 + M^2} \\
\omega_f^2 &=& \frac{\alpha + \sqrt{\alpha^2 - 4 m^2 M^2}}{2} \approx \frac{B^2}{f^2}+M^2 \\
\alpha &=& \frac{B^2}{f^2} + m^2 + M^2
\eea
where the approximate sign shows the small $m$ limit.  Including friction as a perturbation we find that the effective damping terms are
\bea
\Gamma_s &\approx& \frac{1}{B^2/f^2 + M^2 + m^2 } \left [ \Gamma_\phi (M^2 - \omega_s^2 ) + \Gamma_{A_d} (m^2 - \omega_s^2 )\right ]  \nonumber \\
\Gamma_f &\approx& \Gamma_\phi + \Gamma_{A_d}.\nonumber
\eea
This shows that the effect of mixing is to decrease (increase) both the oscillation frequency and the friction depending on which of the modes is being examined. The change is only modest if $M > B/f$ but can be substantial for $B/f \gg M$.

Solving for the slow mode we find
\begin{equation}
	A_{d,x} = - i \frac{\omega_s B/f}{M^2 - \omega_s^2} \phi \approx - i \frac{m}{M} \frac{B/f}{\sqrt{M^2+B^2/f^2}} \, \phi.
\end{equation}
From this one sees that if $m \ll M$, the slow mode is approximately pure scalar field. Therefore, if at early times only the scalar field is displaced from its minimum, only the slow mode is significantly excited.

Furthermore, in the $B/f \gg M$ regime, there is almost no kinetic energy associated with the slow mode because the oscillation frequency is highly suppressed. In fact, instead of having an energy density that oscillates between being purely potential energy to purely kinetic energy, the large mixing causes the energy of the slow mode to oscillate between the scalar potential energy, $m^2 \phi^2$ and the vector potential energy $M^2 A_{d,x}^2$.

Note that even though the equations in this section can be solved exactly, in the small $m$ limit they can be greatly simplified by dropping the $\ddot A_{d,x}$ and $\Gamma_{A_d} \dot A_{d,x}$ terms, as they are subdominant to the mass term in Eq.~(\ref{Eq:simple2}).  Ignoring these terms, $A_d$ can be explicitly written in terms of $\phi$ and the equation for $\phi$ becomes much simpler, leading to the small $m$ limit results very directly. This trick will prove very useful in the next section when the equations cannot be solved exactly.

\section{Gliding in an expanding Universe}

Using the intuition from the previous section, we proceed to study the effects of mixing in the model of Eq.~(\ref{eq:lagrangian-axion}) in an expanding Universe. The equations of motion for the homogeneous fields become
\begin{eqnarray}
\label{Eq: expand}
	\ddot \phi + 3 H \dot \phi + m^2 \phi = \frac{B(t)}{a f} \dot A_d \\
	\ddot A_d + H \dot A_d + M^2 A_d = - \frac{a B(t)}{f} \dot \phi \, ,
\end{eqnarray}
where we assumed $B(t)$ to be a homogeneous background magnetic field (no longer necessarily constant in time), $A_d$ to be the component of the massive gauge field $A_{d,i}$ in the direction of the magnetic field, $H$ the Hubble expansion and $a$ the scale factor.

When $H$ is larger than all other factors $(m, \, M, \, B(t)/f)$ the fields are effectively frozen. Since we are interested in the low frequency mode, and with the expectation from last section that $\omega_s < m, \, M$, we will study the system dynamics when $H < m, \,M$, while $ m \ll B(t)/f$ and $M$.   As mentioned before, for $m \ll M$, the $\ddot A_d$ and $H \dot A_d$ terms are subdominant compared to the mass term of $A_d$ and can be neglected. In this limit the slow mode Eq.~(\ref{Eq: expand}) can be simplified into a second order equation for $\phi$:
\bea
	& &\ddot \phi  + \left ( (3 - 2 \lambda )H + \lambda \frac{\dot B(t)}{B} \right) \dot \phi =  -\frac{m^2 M^2}{B(t)^2/f^2 + M^2} \phi \, , \nonumber  \\
	& & \lambda = \frac{B(t)^2/f^2}{B(t)^2/f^2 + M^2}  . 
\label{eq:effective-second-order}
\eea

Most of the important effects can be directly read from the above equation. The (time dependent) frequency of the mode is the small $m$ limit of $\omega_s$ in Eq.~(\ref{eq:omega_s}). It also explicitly shows how the friction term interpolates between  $3 H$ in the $M \gg B/f$ limit to $H$ in the $B/f \gg M$ limit, which leads to the associated energy density diluting as $a^{-1}$ instead of $a^{-3}$ in the large $B$ regime (we show that the $\dot B(t)/B(t)$ term does not lead to any additional friction effect in what follows).

We can solve Eq.~(\ref{eq:effective-second-order}) in the limit $H < \omega_s$ using the adiabatic approximation:
\begin{equation}
	\phi(t) = \varphi(t) e^{i \int_{t_0}^t dt' \omega_s(t')} \, ,
\end{equation}
where $\varphi(t)$ varies slowly compared to the effective frequency $\omega_s$ (i.e., $|\dot \varphi| \ll \omega_s |\varphi|$ and $|\ddot \varphi| \ll \omega_s |\dot \varphi|$). With this approximation one finds
\begin{equation}
	\varphi(t) = \varphi(t_0) e^{-\frac{1}{2} \int_{t_0}^{t} dt' H \left( 1 + \frac{2 M^2}{M^2+ B^2/f^2} \right)} \, ,
\end{equation}
which satisfies the slow varying condition for $\varphi$ since $H \ll \omega_s$ and shows that the amplitude of $\phi$ scales as $a^{-1/2}$ while $B/f > M$ and as $a^{-3/2}$ when $M > B/f$. It also confirms that the term containing $\dot B$ in Eq.~(\ref{eq:effective-second-order}) does not affect the amplitude of $\phi$.

\begin{figure}
\includegraphics[width=8.5cm,height=5cm]{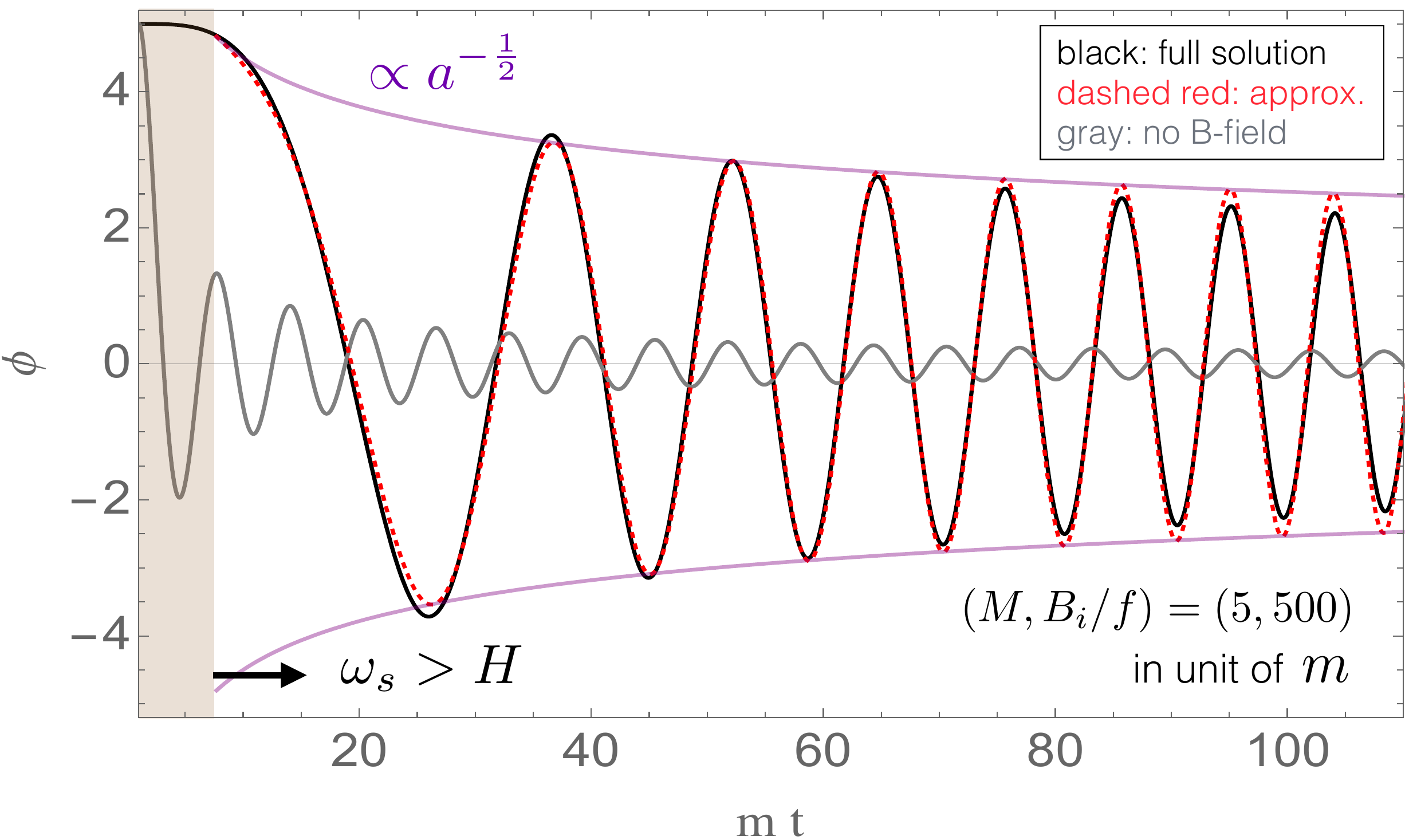}
\caption{A comparison between the full numerical solution of Eq.~(\ref{Eq: expand}) (black) and the approximation from Eq.~(\ref{eq:approx}) using $\omega_s^2\approx m^2M^2/(B(t)^2/f^2+M^2)$ (red dotted). Here $B(t)=B_ia^{-2}$. The gray curve shows the solution of $B_i=0$, which oscillates at a larger frequency and larger damping as compared to the $B_i\neq 0$ case.}\label{fig:plot}
\end{figure}

Let us briefly summarize the cosmological history of the scalar $\phi$ in the limit where $m \ll M$ and $B(t)/f \gg M$ initially.
From Eq.~(\ref{eq:effective-second-order}) we see that the field remains frozen until $\omega(t) \sim H$. Once $H \lesssim \omega(t)$, we can solve the equation in the small $H$ limit and find
\begin{equation}\label{eq:approx}
	\phi(t) \approx \phi_0 \sqrt{\frac{a_0}{a(t)} } \, \exp\left( i \int_{t_0}^{t}  \omega_s(t') dt' \right) \, ,
\end{equation}
where $a_0$ is the scale factor when oscillations start. Eventually $B(t)/f \sim M$ and the approximation of large $B(t)/f$ breaks down.  At this point, the scalar behaves like a normal scalar field, feeling the full effect of Hubble friction, and oscillating at the usual frequency $m$.  Note that this approximation of the cosmological history completely neglects the transition regions where $H \sim \omega(t)$ and $B(t)/f \sim M$, however we have numerically checked the solution and found these approximations to be a reasonable fit as shown in Fig.~\ref{fig:plot}. 

This effect can also be generalized to cases when the masses of $\phi$ and $A_d$ are time dependent. The solutions to the general case are presented in the supplementary materials~\cite{supplement}. 

\section{An axion like particle example}

As an example of how the above mechanism can impact the cosmology of models beyond the standard model, we first show how to increase the number abundance of an axion-like particle (ALP) that obtains its abundance from the misalignment mechanism. A large variety of light ALPs are expected from string theory considerations~\cite{Arvanitaki:2009fg}, which has motivated a rich experimental program searching for their signatures~(see e.g.~\cite{Arias:2012az,Irastorza:2018dyq}). For a given field range of the ALP, there is a direct relation between its mass and its cosmological abundance. We will show that through single derivative mixing such a tight connection can be broken, enlarging the range of parameter space for which ALPs can naturally constitute dark matter.

We will take a FRW universe that is radiation dominated and approximate the ALP potential by its mass term.  In the standard scenario an ALP $\phi$ would begin to oscillate at a temperature $T_{osc}$ when $3 H(T_{osc}) \approx m$.  Before that it behaves as a cosmological constant and afterwards it behaves like cold dark matter.  Once the scalar starts oscillating its comoving number density is approximately conserved and its final abundance (normalized by the photon entropy density) is
\bea
Y = \frac{n_\phi}{s_\gamma} \sim \frac{m \phi_0^2}{T_{osc}^3} \sim \frac{\phi_0^2}{\sqrt{m} M_p^{3/2}} \, ,
\eea
where $\phi_0$ is the initial misalignment of the scalar. This shows that for a given mass and field range, the ALP abundance is approximately fixed (assuming that $\phi_0 \sim f$, where the latter stands for the ALP field range).

We now modify the situation by utilizing a $B$-field combined with mixing as in the previous section.  We will take the $A_{\mu}$ to have a homogeneous $B$-field that scales as $B(T) = \beta T^2$ and for simplicity we will take $M \gg m$ and assume that $B/f \gg M$ when oscillations begin.  Because the frequency of oscillation is greatly suppressed by the mixing, the scalar will now start to oscillate at a lower temperature $T'_{osc}$ when $H(T'_{osc}) \approx \omega \approx m M f/B(T'_{osc})$. This effect makes the energy density in the scalar remain constant for a longer time. In addition, Hubble friction is effectively reduced in this limit so that the ALP energy density only dilutes away as $a^{-1}$ after oscillations start. Both effects enhance the final abundance of $\phi$. Normal Hubble friction turns on once $B(T_B)/f \sim M$, and from this point on the scalar dilutes away like cold matter.

After accounting for these additional effects, the final abundance of the scalar is
\bea
Y \sim \frac{m \phi_0^2}{T_B^2 T'_{osc}} \sim \frac{m^{3/4} \phi_0^2 \beta^{5/4}}{f^{5/4} M^{5/4} M_p^{1/4}}
\eea
Giving a total enhancement factor of
\bea
\frac{Y(\beta)}{Y(\beta=0)} \sim \left ( \frac{m M_p \beta}{M f}\right )^{5/4} \sim \left ( \frac{T_{osc}}{T'_{osc}} \right )^5 \, .
\eea
So that as long as the initial $B$-field is sufficiently large, the abundance of the scalar can rise drastically.

\section{Enhancing the QCD axion number abundance}

The QCD axion is one of the best motivated dark matter candidates as it can solve the strong CP problem~\cite{Peccei:1977hh,Peccei:1977ur,Weinberg:1977ma,Wilczek:1977pj} and be dark matter at the same time~\cite{Abbott:1982af,Dine:1982ah,Preskill:1982cy}.  Unlike an ALP, the axion mass and field range are tied together, making its final number abundance only function of its decay constant $f_a$.  The correct dark matter abundance is only obtained when $f_a \sim 10^{11} - 10^{12}$ GeV in the minimal setup.  There have been several attempts to increase the abundance of the QCD axion for the smaller $f_a$ values so that it can play the roll of dark matter even in the small $f_a$ limit~\cite{Turner:1985si,Lyth:1991ub,Visinelli:2009kt,Hiramatsu:2012sc,Co:2017mop,Co:2018mho,Co:2019jts,Chang:2019tvx,Harigaya:2019qnl}. In this section we show that our gliding scalar mechanism provides a simple mechanism for this and can extend the parameter space ($10^9$ GeV $\lesssim f_a \lesssim 10^{12}$ GeV) over which the QCD axion can be dark matter.

The QCD axion Lagrangian that we will consider is
\bea\label{eq:aQCD}
\mathcal{L} =  \mathcal{L}_\text{kinetic} + \frac{M^2}{2} A_d^\mu A_{d,\mu} + \frac{\phi}{2 f_\gamma} F_d \tilde F + \frac{\alpha}{8 \pi} \frac{\phi}{f_a} G \tilde G.
\eea
The only difference between the QCD axion and the ALP is that due to the gluon $G \tilde G$ coupling, its mass has temperature dependence.  In the following section, we approximate the axion mass by a simple time-dependent form
\bea
m_a(T) = m_a(T=0) \times \left\{
	\begin{array}{ll}
		\left( \frac{T_0}{T} \right)^4 \, , & \text{for } T>T_0 \, , \\
		1 \, , 	& \text{for } T \leq T_0
	\end{array}	\right.
\eea
where $T_0 \sim 100$ MeV~\cite{Gross:1980br}. As with the ALP, the axion starts to oscillate when
\bea
3 H(T_{osc}) \approx m_a(T_{osc}) \, ,
\eea
afterwards the axion number density, $m_a(T) \phi^2$, dilutes away as $a^{-3}$. Note that while the mass is increasing, comoving axion number density conservation implies that the axion amplitude is changing faster than $a^{-3/2}$.  The end result is an abundance
\bea
Y \sim \frac{\phi_0^2}{T_{osc} M_p} \, .
\eea

In the presence of a $B$-field and a heavy dark photon $A_d$, with $M > m_a$, the axion number abundance from misalignment  can be enhanced.  As before, we take the $B$-field to be $B = \beta T^2$.  The axion starts to oscillate when
\bea
H(T_{osc}') = \omega_s(T_{osc}') \approx \frac{m_a(T_{osc}') M f_{\gamma}}{B(T_{osc}')}\,.
\eea
Afterwards its evolution is well described by (see supplementary materials for details~\cite{supplement})
\bea
\phi(t) &\approx& \phi_0 \sqrt{\frac{m_a(T_{osc}')}{m_a(T)}\frac{T}{T_{osc}'}} \, \exp\left( i \int  \omega(t) dt \right) \, ,
\eea
until the magnetic field has decayed sufficiently such that
\bea
\frac{B(T_{f})}{f_{\gamma}} = M .
\eea
Finally, after this point, the axion dilutes away as normal.  The final number abundance of the axion is
\bea
Y \sim \frac{m_a \phi_0^2 T_0^4}{T_f^2 T_{osc}'^5}\,,
\eea
which gets an enhancement comparing to the case with no $B$-field up to a factor
\bea
\frac{Y(\beta)}{Y(\beta=0)} \sim \left ( \frac{T_{osc}}{T'_{osc}} \right )^{13} \lesssim \left(\frac{M_p \,f_a^2}{f_{\gamma}^3}\right)^{\frac{13}{24}}\,.
\eea
We have used the fact that the enhancement is maximized when $M \sim m_a \sim T_0^2/f_a$.  This leads to an enhancement factor that allows for QCD axion dark matter to come entirely from the misalignment mechanism all the way down to where supernova constraints become important, $f_a \sim 10^9$ GeV~\cite{Raffelt:1996wa,Chang:2018rso}.  More detail about this particular scenario will be given in Ref.~\cite{longpaper}.

\section{Inflating on a non-slow roll potential}

In this section, we show that small magnetic fields can even allow for a short period of inflation on potentials that normally do not support slow roll inflation.
The Lagrangian we will consider is
\bea
\mathcal{L} =  \mathcal{L}_\text{kinetic} - \frac{m^2}{2} \phi^2 + \frac{M^2}{2} A^\mu_d A_{d,\mu}  + \frac{\phi}{2 f} F_d \tilde F .
\eea
We will take as initial conditions, $\phi \approx \phi_0$, $\dot \phi \approx 0$ and $B_0^2 \ll m^2 \phi_0^2$ as well as taking $M > m$.  The energy density of the universe is initially dominantly in $\phi$, with a small component in the magnetic field.  To illustrate the effect, we will take $m > H$ or equivalently $\phi_0 < M_p$, so that this potential does not support slow roll inflation.

The scalar $\phi$ oscillates with a frequency $\omega_s(t) = m M f/B(t) \ll m$ so it behaves as a cosmological constant until $H \sim \omega_s(t)$, after which inflation ends.  We will be concerned with how many e-foldings can inflation last.  Inflation ends when $B_f/f \sim M$, where we take $B_0$ to be the initial $B$-field and $B_f$ to be the final $B$-field and assume the usual field dilution $B \propto a^{-2}$.
\bea
N_e = \frac{1}{2} \log \frac{B_0}{B_f} \sim \frac{1}{2} \log \frac{c \, m \phi_0}{M f} < \frac{1}{2} \log \frac{c \,\phi_0}{f} \,
\eea
where $c<1$ encodes the initial fraction of the energy density of the universe stored in the magnetic field, $B_0 = c \, m \, \phi_0$.
To maximize the number of e-foldings, we expect $M \sim m$ and $\phi_0 \gg f$.
For the EFT to be consistent, we require
\bea
m^2 \phi_0^2 \sim T_{RH}^4 < f^4
\eea
where $T_{RH}$ is the reheat temperature of the universe assuming instantaneous reheating.

Thus we see that we have the bound on the number of e-foldings of inflation to be 
\bea
N_e < \frac{1}{2} \log \frac{c \, M_p}{T_{RH}}  \lesssim 23
\eea
where we have required that the reheat temperature is above an MeV and that the fraction of initial energy density stored in the magnetic field is $\mathcal{O}(1\%)$.  Thus we find the amusing result that a $B$-field can cause 20 odd e-foldings of inflation on a potential that normally does not support slow roll inflation.  
At the end of inflation, the energy in the magnetic field is $10^{40}$ times smaller than the energy in the inflaton, demonstrating that even an extremely tiny magnetic field can have rather drastic consequences on the evolution of the inflaton.
It would be interesting if there were models in which all of inflation could be due to this effect, which could potentially lead to interesting observational consequences due to anisotropic contributions due to the magnetic field~\cite{Maleknejad:2012fw}.

\section{A completely friction-less scalar}

In the scenarios we studied so far, the energy density oscillates between the potential energy of a scalar field and the potential energy of a massive vector field.  Since the energy density of the vector field redshifts even when the field amplitude $A_d$ does not decay, the energy of the coupled system still dilutes with the Hubble expansion. In this section we entertain the idea of getting a completely friction-less system by focusing on a scenario with only scalars. 

In analogy with the previous example, one could construct a five scalar model with a mixing term $\phi_1 \partial_\mu \phi_2 \partial_\nu \phi_3 \partial_\rho \phi_4 \partial_\sigma \phi_5 \epsilon^{\mu \nu \rho \sigma}$. If three scalars have non-trivial spatially dependent backgrounds, the end result is a scalar that does not dilute away in time.  Rather than studying this complicated and admittedly implausible example, we consider instead a three scalar system with a mixing term
\bea
\mathcal{L} \supset - \frac{m^2}{2} \phi^2 - \frac{M^2}{2} \Phi^2 + \frac{\phi}{f} \partial_\mu \Phi \partial^\mu \Psi
\eea
Instead of a $B$-field, we will take $\Psi$ to have a large kinetic term (velocity).  In order to make this two time derivative mixing act like a single time derivative mixing, we take $\dot \Psi$ to dominate the energy density of the universe so that we can treat it as a background that dilutes away as $\dot \Psi \propto a^{-3}$.  The equations of motion are
\bea
\ddot \phi + 3 H \dot \phi &=& - m^2 \phi + \frac{\dot \Psi}{f} \dot \Phi\,, \\
\ddot \Phi + 3 H \dot \Phi &=& - M^2 \Phi - \frac{\dot \Psi}{f} \dot \phi \,.
\eea
It can be easily shown that we can neglect the effect of $\phi$ and $\Phi$ in the dynamics of $\Psi$ as long as $\dot \Psi^2 \gg m^2 \phi_0^2$. Following the analysis done in the previous section, it is easy to see that the scalar $\phi$ does not dilute away in time as long as $\dot \Psi / f \gtrsim m , \, M$ and $\dot \Psi^2 \gg m^2 \phi_0^2$.
\\
\\
\section{Conclusion}

In this article, we highlighted the effects of single derivative mixing between particles, as opposed to the usual mass mixing or kinetic mixing. In the regime of large mixing, the time evolution of the system is dictated by a single time derivative mixing term, similar to what happens in slow-roll inflation where the dynamics is dictated by the single time derivative Hubble friction term.  In both of these cases, the evolution of the scalar and its associated energy density is drastically modified.
The single derivative interaction we considered has two effects.  The first is that it leads to a significant reduction of the oscillation frequency for massive scalars, somewhat similar to how Hubble friction can completely prevent the oscillation of a scalar.  However because our single derivative term is a mixing term rather than a friction, the total energy density in the system oscillates between the potential energy of two fields with negligible contributions to their respective kinetic energies.  This has the unique features of reducing the frictional effects and delaying the onset of oscillations.

These effects present a myriad of interesting opportunities, only a few of which were briefly explored in this article.
Utilizing $B$-fields, the number abundance of QCD axions could be enhanced allowing for the axion to be dark matter for $f_a$ all the way down to the super nova bound, a notoriously difficult feat.
Single derivative mixing with $B$-fields may also have interesting applications during inflation, as it can support short periods of inflation on potentials that do not normally allow for inflation.
It will be exciting to see what opportunities single derivative mixing bring.

\section*{Acknowledgements}
The authors thank Michael Fedderke and Junwu Huang for useful conversations. This research was supported in part by the NSF under Grant No. PHY-1914480, PHY-1914731 and by the Maryland Center for Fundamental Physics (MCFP).  
YT was also supported in part by the US-Israeli BSF grant 2018236. 
The authors also thank the KITP institute (Enervac19 program), Aspen Center for Physics, and the Munich Institute for Astro- and Particle Physics (MIAPP) of the DFG Excellence Cluster Origins, where part of this work was conducted, for hospitality.

\bibliography{reference}


\clearpage
\newpage
\maketitle
\onecolumngrid
\begin{center}
	\textbf{\large Scalars gliding through an expanding Universe} \\ 
	\vspace{0.05in}
	{ \it \large Supplementary Material}\\ 
	\vspace{0.05in}
	{}
	{Anson Hook, Gustavo Marques-Tavares, and Yuhsin Tsai}
	
\end{center}
\setcounter{equation}{0}
\setcounter{figure}{0}
\setcounter{table}{0}
\setcounter{section}{0}
\renewcommand{\theequation}{S\arabic{equation}}
\renewcommand{\thefigure}{S\arabic{figure}}
\renewcommand{\thetable}{S\arabic{table}}
\newcommand\ptwiddle[1]{\mathord{\mathop{#1}\limits^{\scriptscriptstyle(\sim)}}}

\renewcommand{\theHequation}{Supplement.\theequation}
\renewcommand{\theHtable}{Supplement.\thetable}
\renewcommand{\theHfigure}{Supplement.\thefigure}

\section{Time varying masses}

In this section, we show how the field evolution changes when the masses of the particles depend on time. Considering the case where both masses could have time dependence changes Eq.~(10) to
\begin{equation}
\ddot \phi  + 3 H \dot \phi + m(t)^2 \phi =  - \frac{B(t)}{a f} \frac{d}{dt} \left [ \frac{a B(t) \dot \phi}{M(t)^2 f}\right ]\,
\end{equation}
Using the WKB approximation we write
\begin{equation}
\phi(t) = \varphi(t) e^{i \int_0^t dt^\prime \omega_s(t^\prime)} \, ,
\end{equation}
where we assume $\varphi(t)$ is varying slowly, i.e. $\dot \varphi \ll \omega_s \varphi$ (and also $\ddot \varphi \ll \omega_s \dot \varphi$). Under these approximations the equation for $\varphi$ becomes
\begin{equation}
\dot \varphi = - \frac{1}{2}\left( (3 - 2 \lambda) H - \lambda \frac{\dot M}{M} + \frac{\dot m}{m} \right) \varphi \, ,
\end{equation}
where $\lambda(t) = \frac{B^2(t)/f^2}{B^2(t)/f^2 + M^2(t)}$.
While the above equations can be solved, the form tends not to be particularly enlightening as it involves integrals over the time dependent functions $(B(t), \, M(t), \, m(t))$.  To simplify matters, we will take the magnetic field and the mass term to take the form
\bea
M(t) = M_0 \left ( \frac{t}{t_0} \right)^a \qquad B(t) = B_0 \left ( \frac{t_0}{t} \right)^b
\eea
while $m(t)$ can remain arbitrary (as long as the WKB approximation remains valid, i.e. $\dot m \ll \omega_s m$).  $\varphi$ can now be solved for to obtain
\bea
\label{eq:mass-changing}
\phi(t) &\approx& \phi_0 \sqrt{\frac{m_0}{m(t)} \frac{M(t)}{M_0} \frac{a_0}{a(t)}} \left( \frac{B(t)^2/f^2}{B(t)^2/f^2 + M(t)^2} \frac{B_0^2/f^2 + M_0^2}{B_0^2/f^2} \right )^{\frac{a+1}{4 (a + b)}}  \, \exp\left( i \int  \omega_s(t) dt \right) \, .
\eea

In the limit of $B/f \gg M$, $\phi$ dilutes away with a factor of $1/\sqrt{m(t)}$ as expected from number density conservation, $1/\sqrt{a(t)}$ corresponding to the reduced Hubble friction, and $\sqrt{M(t)}$ due to mixing with the gauge boson.
In the limit of $B/f \ll M$, $\phi$ dilutes away like a normal scalar field with no mixing.

Interestingly, it can be easily shown that a very good approximation of Eq.~(\ref{eq:mass-changing}), is to treat the scalar as diluting away as $\sqrt{\frac{m_0}{m(t)} \frac{M(t)}{M_0} \frac{a_0}{a(t)}}$ until $B(t_c)/f = M(t_c)$, and then afterwards as a normal scalar $\sqrt{\frac{m(t_c)}{m(t)}} \left(\frac{a(t_c)}{a(t)}\right)^{3/2}$.  This justifies the approximations made in the main text regarding the QCD axion abundance.

\section{Non Homogenous $B$-fields}

In this section, we generalize our previous results to the case of a non-homogenous $B$-field. For simplicity we will approximate the $B$-field by a single Fourier mode in a fixed direction: 
\begin{equation}
\label{eq-Bx}
\vec B (x,t) = B(t) \cos (kx) \hat n \, ,
\end{equation} 
with $\hat x \cdot \hat n = 0$. The equations of motion for the system (treating the magnetic field as a background field) are:
\begin{equation}
\begin{aligned}
\ddot \phi + 3 H \dot \phi - a^{-2}\partial_i^2 \phi + m^2 \phi - \frac{1}{2 a^3} \epsilon^{\mu \nu i j} \partial_\mu A_{d, \, \nu} F_{ij} &= 0 \, , \\
\partial_i F_{d, \, i0} - \frac{1}{2 a} \epsilon^{0ijk} \partial_i \phi F_{jk} - a^2 M^2 A_{d, \, 0}^2 &= 0 \, , \\
\partial_0 F_{d, \, 0i} + H F_{d, \, 0i} - a^{-2} \partial_j F_{d, \, ji} + \frac{1}{2 a}\epsilon^{0ijk} \dot \phi F_{jk} + M^2 A_{d, \, i} &= 0 \, .
\end{aligned}
\label{Eq:non-homogeneous}	
\end{equation}
Assuming $\hat n$ is constant and that the initial conditions are such that $A_{d \, , \mu} = 0$, one can show that $A_{d, \, 0} = 0$, that only $\vec A_d \propto \hat n$ gets generated and that $\nabla \cdot \vec A_d = 0$. The initial conditions also ensure that the only non-trivial spatial dependence is in the $\hat x$ direction.

\subsection{Analytic results in static universe}

We first study this system analytically in a static universe with a constant (in time) magnetic field. Given the simple spatial dependence for the magnetic field, it is best to work in Fourier space. We will work with the Fourier transforms of $A_d$ and $\phi$.  Since only Fourier modes with momentum equal to integer multiples of $k$ will be excited, we will label the different Fourier modes by $A_j(=A_{d,j}$ from simplicity) and $\phi_j$ where the momentum is $p = j k$.
The equations of motion we have to solve after neglecting the double time derivatives and single time derivatives (except the ones multiplied by $B$) in Eqs.~(\ref{Eq:non-homogeneous}) are
\bea
0 = - (m^2 + j^2 k^2) \phi_j + \frac{i \omega_s B}{2 f} \left( A_{j-1} + A_{j+1} \right)\,,  \\
0 = - (M^2 + j^2 k^2) A_j - \frac{i \omega_s B}{2 f} \left( \phi_{j-1} + \phi_{j+1} \right)\,,
\eea
where we let $j$ go negative and $\phi_j = \phi_{-j}$ ($A_j = A_{-j}$).  From these equations, one can see that $\phi_j$ is excited only for even $j$ while $A_j$ is excited only for odd $j$.
We will solve this set of infinite differential equations for the slow oscillating mode.  In the limit where $k$ is small compared to all scales, the magnetic field appears constant locally and we recover the results presented in the main text.  Thus, we will consider the opposite limit where $k$ is large.

In the large $k$ limit, these equations are
\bea
\phi_j = \alpha_j (A_{j-1} + A_{j+1}),   \qquad
A_j = - \alpha_j (\phi_{j-1} + \phi_{j+1}),\qquad
\alpha_j = \frac{i \omega_s B}{2 f j^2 k^2}, \qquad \alpha_0 = \frac{i \omega_s B}{2 f m^2}\,.
\eea
These equations can be simplified by taking a variable $\vec a_j = (A,\phi)_j$ using the odd or even $j$ to differentiate between $\phi$ and $A$.  One can formally rewrite this equations in terms of a simple recursion relationship:
\bea
\vec a_{j+1} = (-1)^{j+1} \alpha_{j+1} \gamma_{j+1} \vec a_j\,, \qquad \gamma_j = \frac{1}{1+\alpha_j \alpha_{j+1} \gamma_{j+1}}\,.
\eea

In order to obtain analytic expressions, we will work to leading order in $(m/k)\ll1$.  We will end up finding that $\alpha_j \sim m/k$ and $\vec a_j \sim (m/k)^j$.  Thus we will assume these scalings and find a self-consistent solution.  Under these approximations, the equations simplify as
\bea
\gamma_j &=& \frac{1}{1+\alpha_j \alpha_{j+1} \gamma_{j+1}} \approx 1 - \alpha_j \alpha_{j+1} \gamma_{j+1} \approx 1- \alpha_j \alpha_{j+1} + \mathcal{O}(\frac{m^3}{k^3})\,, \\
\vec a_{j+1} &\approx& (-1)^{j+1} \alpha_{j+1} \vec a_j\,.
\eea
We are now in a position to solve for the frequency of oscillation and its corresponding eigenvector.

The frequency can be found by considering the $\vec a_0$ boundary condition
\bea
\phi_0 = 2 \alpha_0 A_1 = - 2 \alpha_0 \alpha_1 \gamma_1 \phi_0 .
\eea
This can be solved to obtain the frequency
\bea\label{eq:freq}
\omega_s^2 = 2 \frac{m^2 k^2 f^2}{B^2} \left (1 - \frac{m^2}{8 k^2}   + \mathcal{O}(\frac{m^3}{k^3}) \right ).
\eea

Similarly the eigenvectors can be calculated, and the general solution for $\phi_j$ and $A_j$ is
\bea
\phi_j &=& \phi_0 \left( \frac{m}{\sqrt{2} k} \right)^{j} \frac{1}{(j!)^2}   + \mathcal{O}((\frac{m}{k})^{j+1})\,, \\
A_j &=& - i \phi_0 \left( \frac{m}{\sqrt{2} k} \right)^{j} \frac{1}{(j!)^2} + \mathcal{O}((\frac{m}{k})^{j+1}) \qquad \forall j \ge 3\,.
\eea
Note that while this approximation was done for small $(m/k)$, it remains numerically a good fit even when $m=k$ because of the $j^{-2}$ factors that appear in all of the expressions.  Therefore the approximate solution works well in the large $j$ approximation as well. What we learn from the results is that even if the $B$-field has a smaller coherence length as compared to the horizon size, we can still focus on a small number of $k$-modes with a modified oscillation frequency in Eq.~(\ref{eq:freq}). 

The biggest upshot of this analytical calculation is that when $k \gg m$ we can focus on only two states, the zero mode of $\phi$ and the first $k$-mode of $A_d$, which leads to an almost identical analysis to the previous section with the identification of the time-dependent mass $M$ by $k/a$.  In fact, as will be shown in the following section, approximating the large $k$ case as a two state system with the vector having a mass $\sqrt{M^2+k^2/a^2}$ will be a very good approximation to the full numerical results.

\subsection{Numerical solution for general case}

\begin{figure}[ht]
	\includegraphics[width=0.32\textwidth]{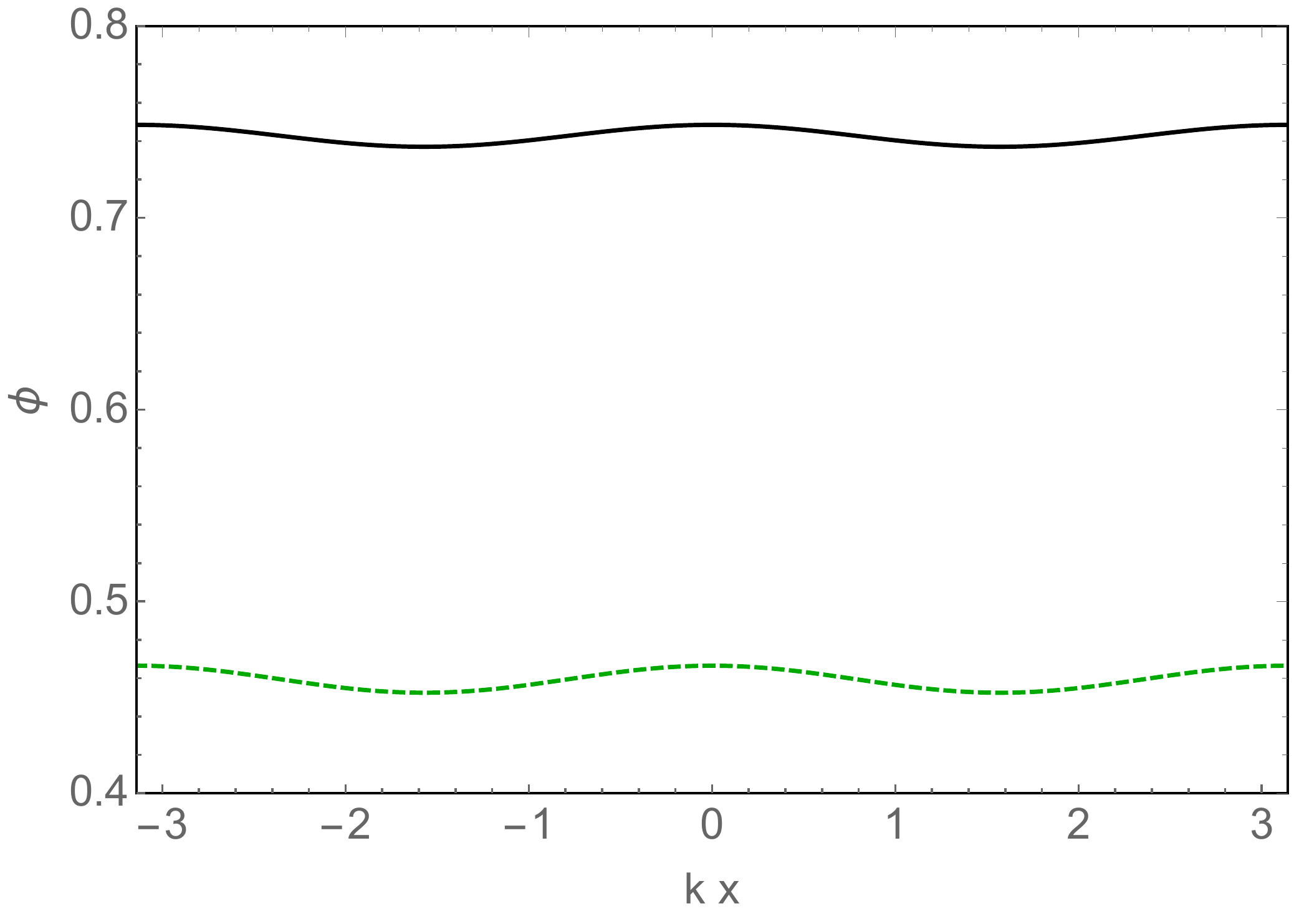}
	\includegraphics[width=0.32\textwidth]{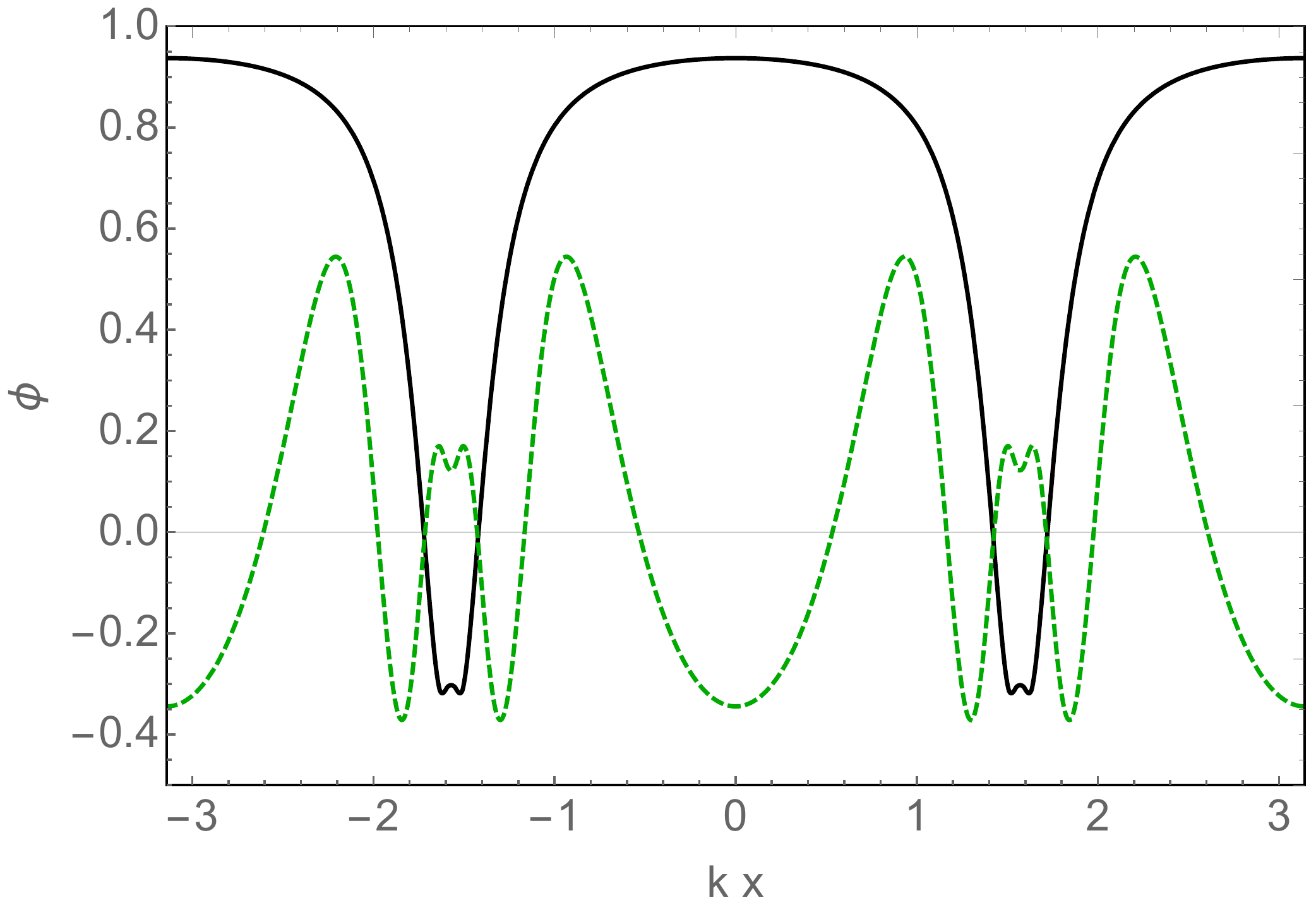}
	\includegraphics[width=0.32\textwidth]{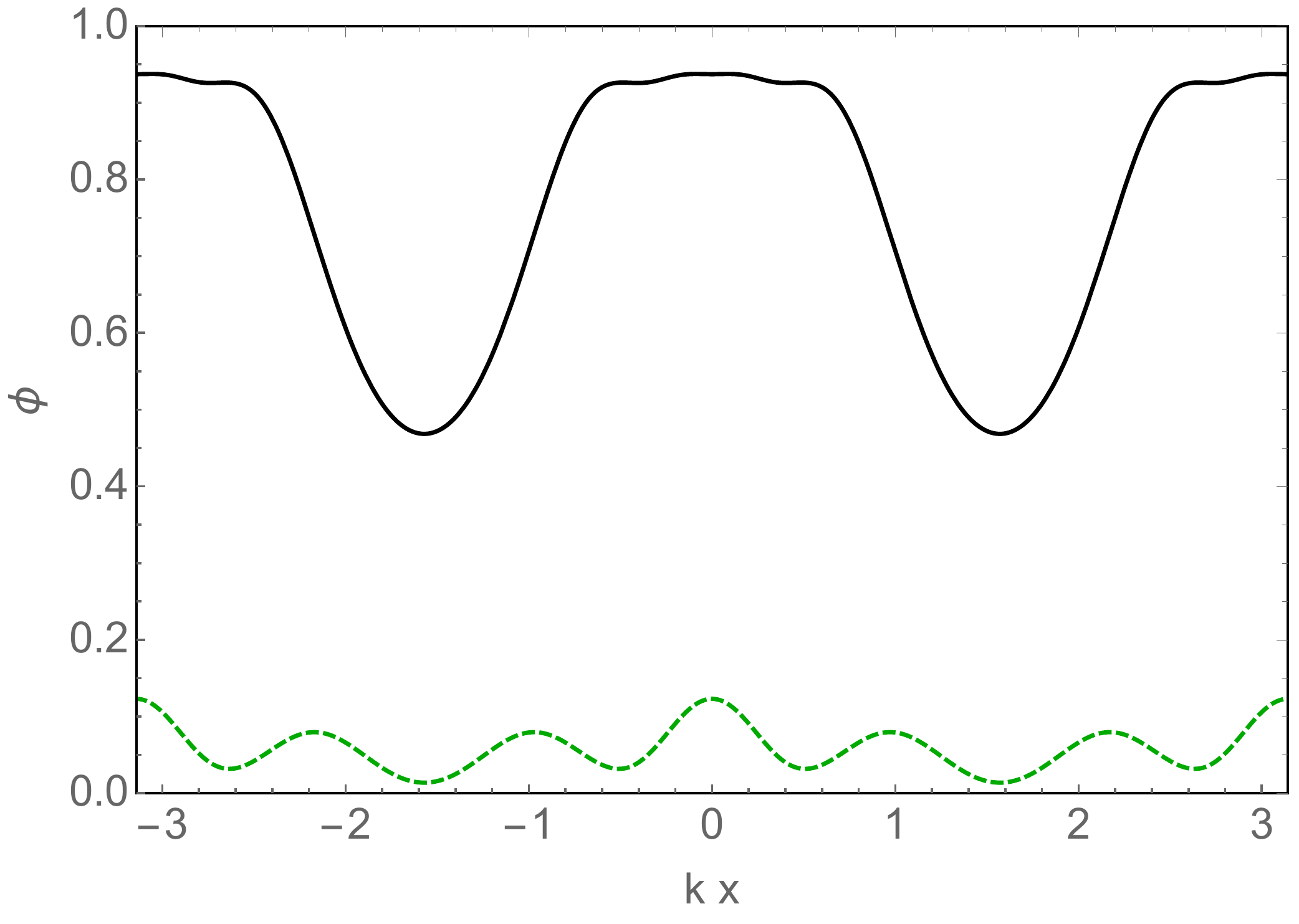}
	\caption{Spatial profile for the numerical solution to $\phi(t,x)$ for two different times, $m t = 4$ (black solid curve) and $m t = 14$ (green dashed curve). The three plots represent different choices for $k$ in Eq.~\ref{eq-Bx}  : {\bf left:} large $k$, ($k/m = 100, \, k/M = 20$); {\bf center:} small $k$, ($k/m = 0.05, \, k/M = 0.01$); and {\bf right:} intermediate $k$, ($k/m = 10, \, k/M = 2$). In all plots we set the scale factor $a_0 = 1$ at $m t_0 = 0.02$ and chose $B_0/f = 2000 M$.}
	\label{fig:spatial-profile}
\end{figure}

\begin{figure}[ht]
	\includegraphics[width=0.32\textwidth]{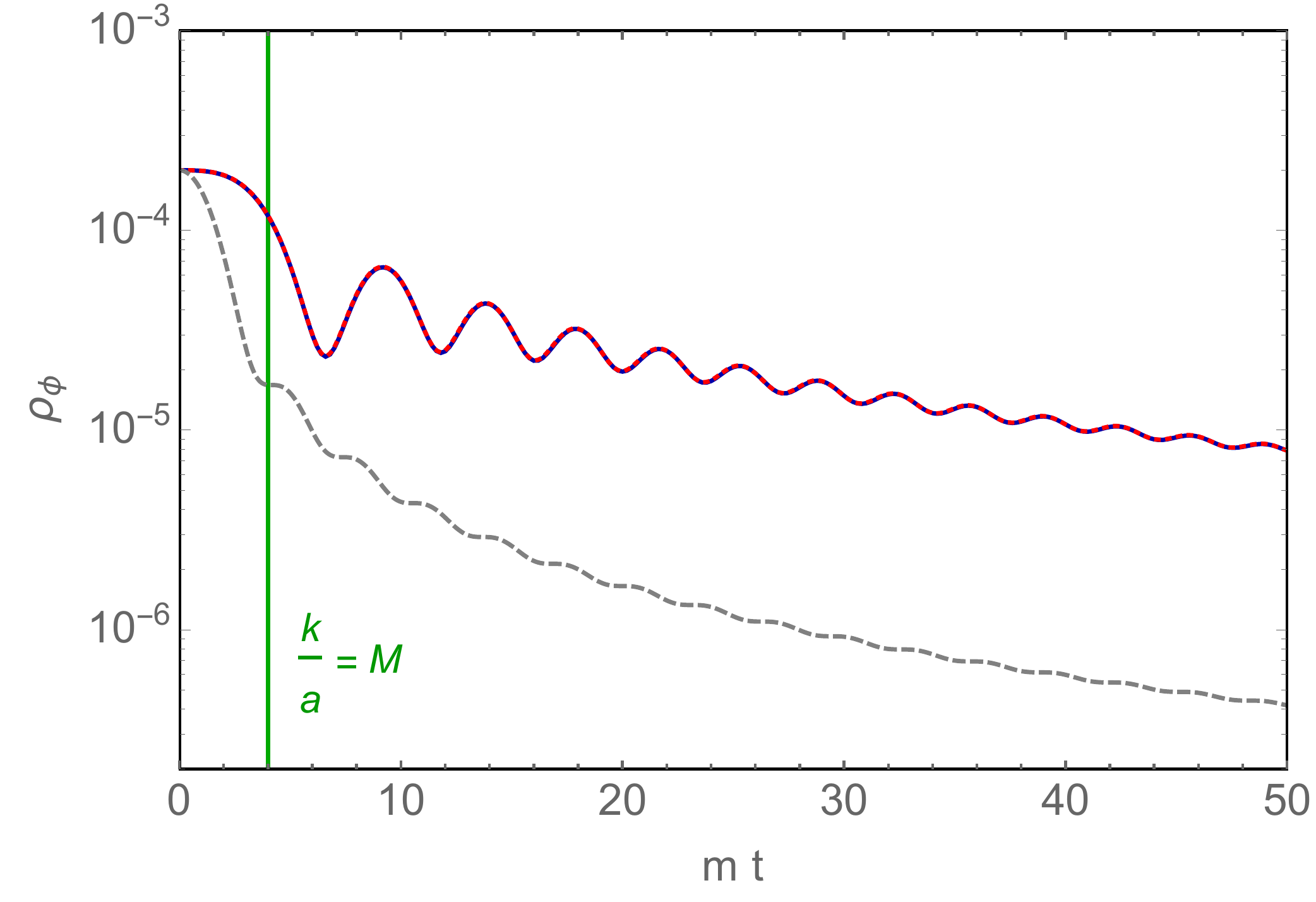}
	\includegraphics[width=0.32\textwidth]{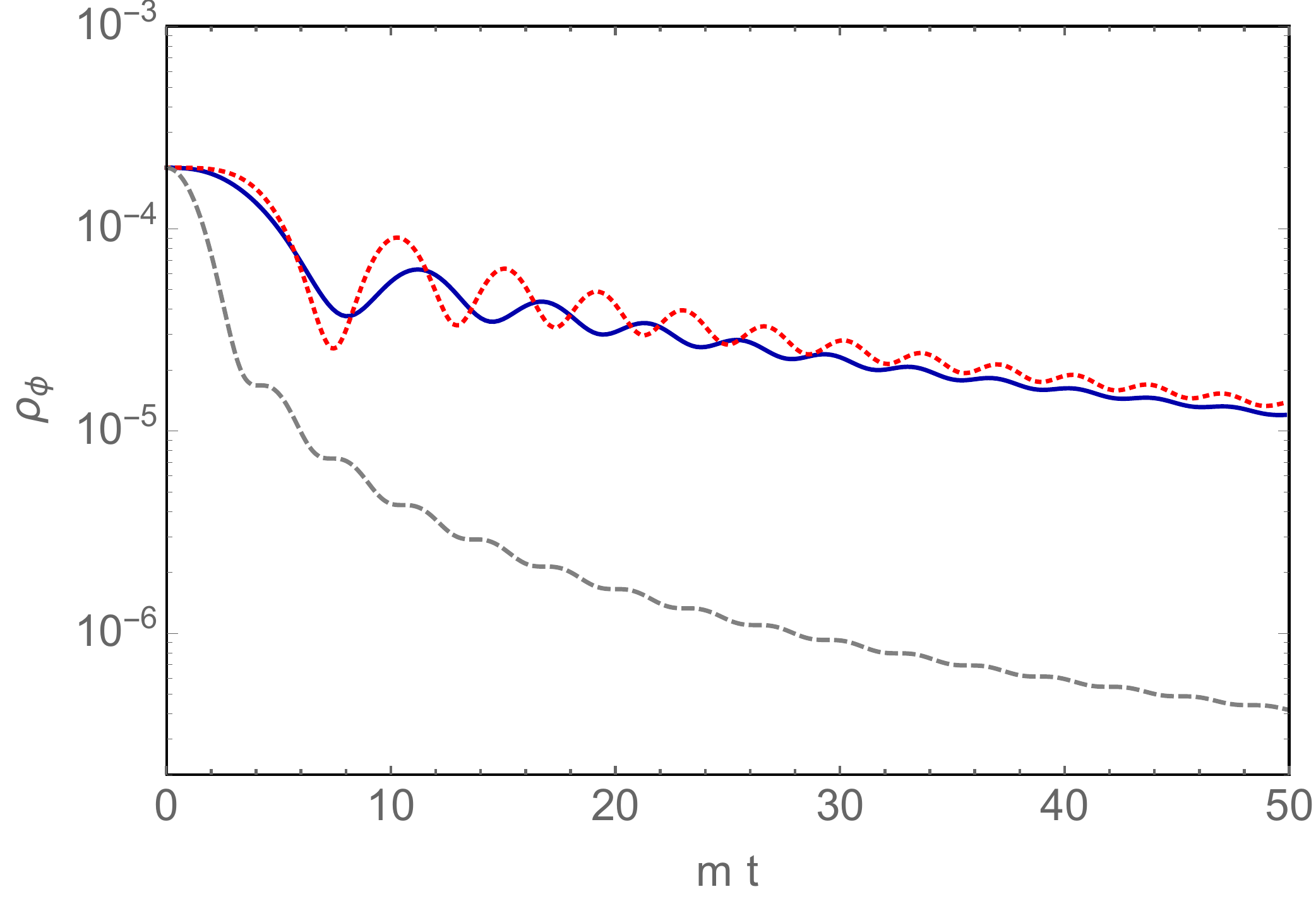}
	\includegraphics[width=0.32\textwidth]{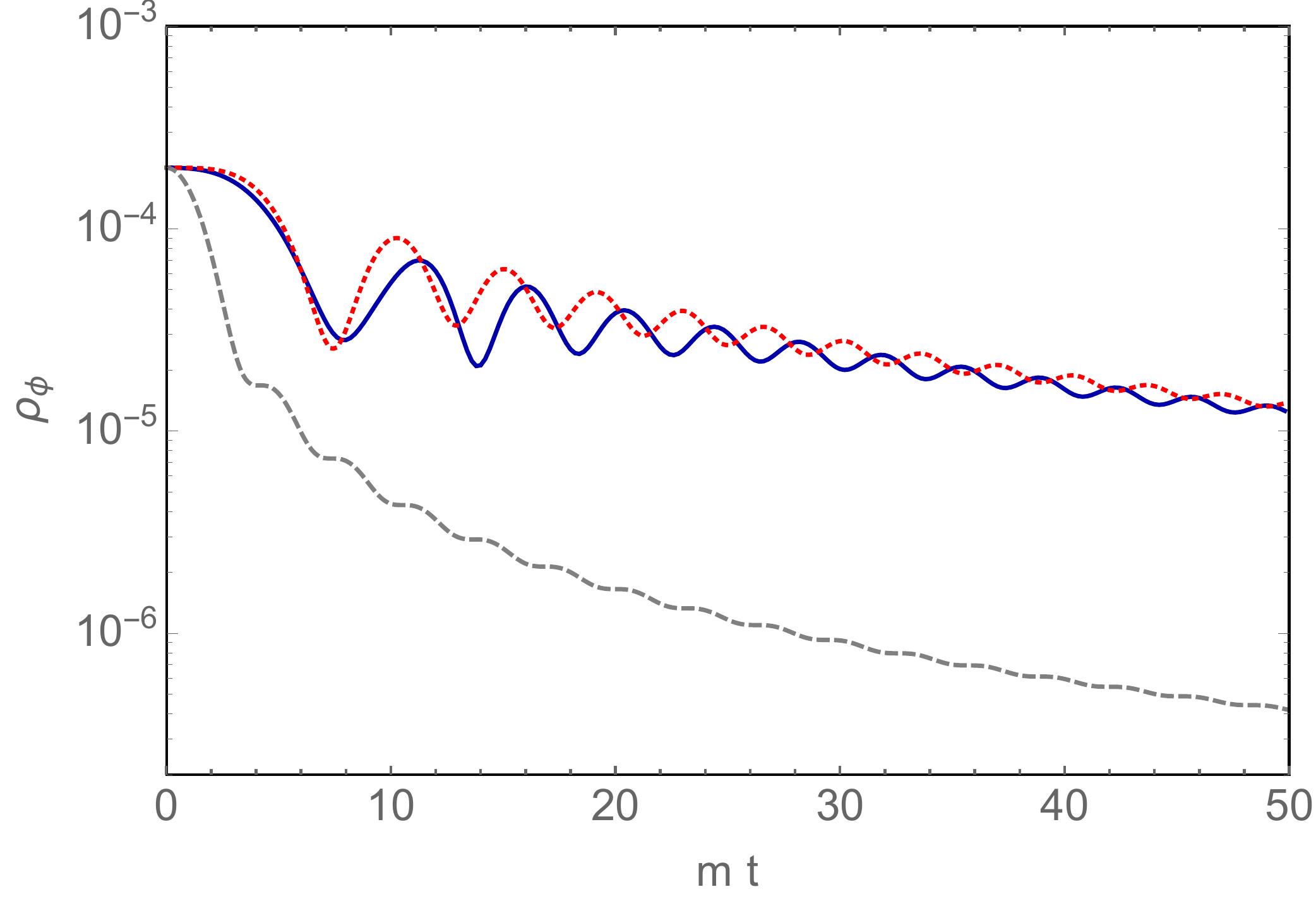}
	\caption{The solid blue curve shows the averaged energy density in the axion field as a function of time in the same three scenarios described in Fig.~\ref{fig:spatial-profile} : {\bf left:} large $k$, ($k/m = 100, \, k/M = 20$); {\bf center:} small $k$, ($k/m = 0.05, \, k/M = 0.01$); and {\bf right:} intermediate $k$, ($k/m = 10, \, k/M = 2$). The dashed gray curve shows the energy density for the homogeneous case with $B = 0$ and the dotted red curve shows the homogeneous case with $B = \frac{B_0}{\sqrt{2}} a^{-2}$ and assuming an effective mass for the dark photon given by $\sqrt{M^2+k^2/a^2}$.}
	\label{fig:rhoa}
\end{figure}

We have also solved Eqs.~\ref{Eq:non-homogeneous} numerically, while imposing periodic boundary conditions for the spatial boundaries. In our numerical results we take $B(t) = B_0 \left(a_0/a \right)^2$, $k(t) = k_0 \left(a/a_0 \right)$ and the Universe to be radiation dominated such that $H \propto a^{-2}$.
Our results are shown in Fig.~\ref{fig:spatial-profile}, where we consider three different scenarios, $k \gg M > m$, $k \ll m < M$ and $m < k \sim M$. In Fig.~\ref{fig:rhoa} we also show the spatially averaged energy density $\rho_\phi$ as a function of time for the different scenarios and compare it with a homogeneous case using an effective mass for the dark photon given by $M_\text{eff} = \sqrt{M^2 + k^2/a^2}$, to compare to the two state approximation discussed in the previous section.

\paragraph{$k \gg M > m$} 
We find that when $k$ is much larger than the other mass scales, the axion remains approximately homogeneous with a small perturbation with wave-vector $2k$.  This result is in accordance with with our analysis in the previous section where we found that the additional frequency dependent pieces of the axion would be at least suppressed by $m^2/k^2$.  As can be seen in Fig.~\ref{fig:rhoa}, the two state approximation suggested by the analytic approximations, is in excellent agreement with the numerical results even after $k/a \lesssim M$.

\begin{figure}[ht]
	\includegraphics[width=0.32\textwidth]{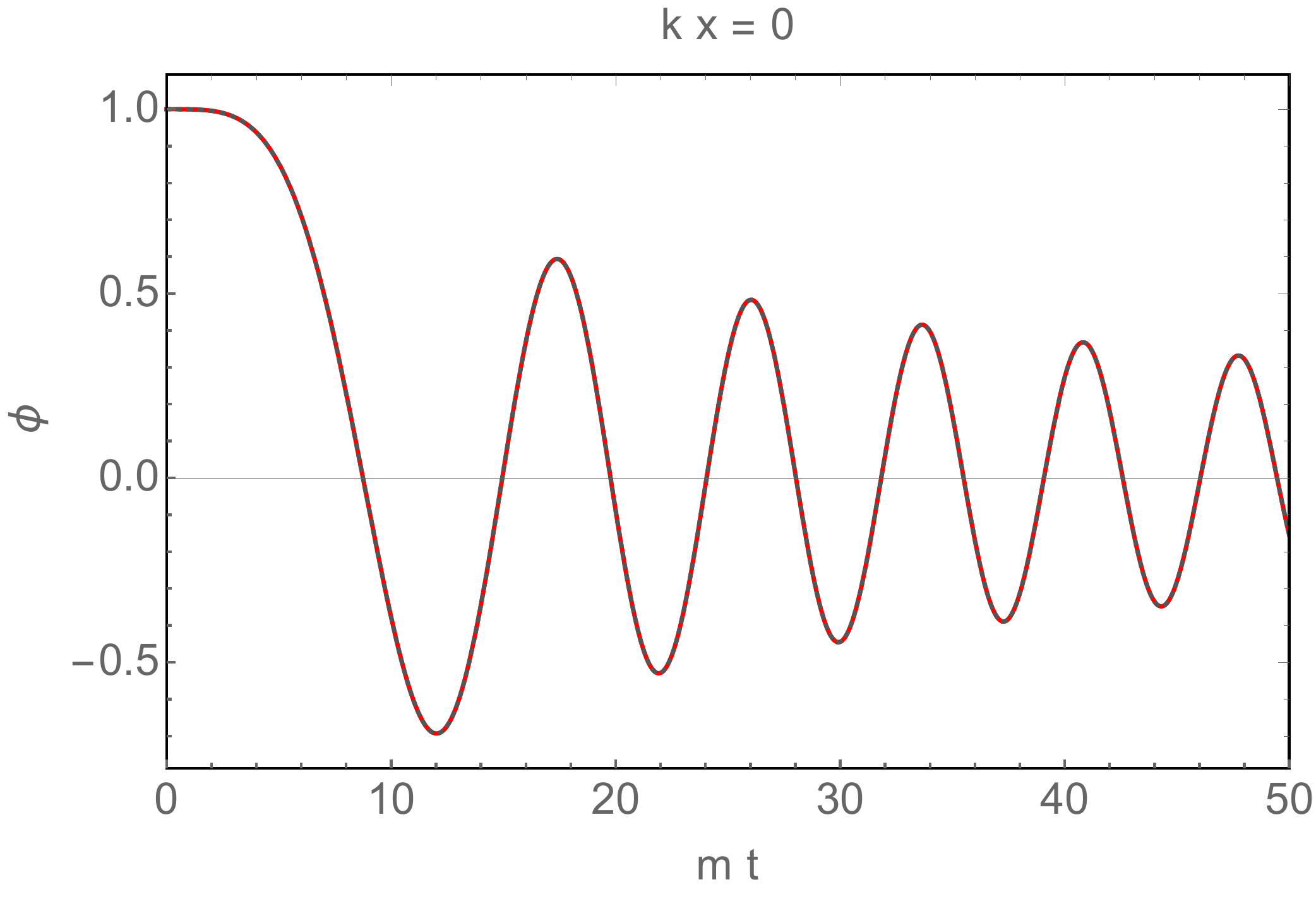}
	\includegraphics[width=0.32\textwidth]{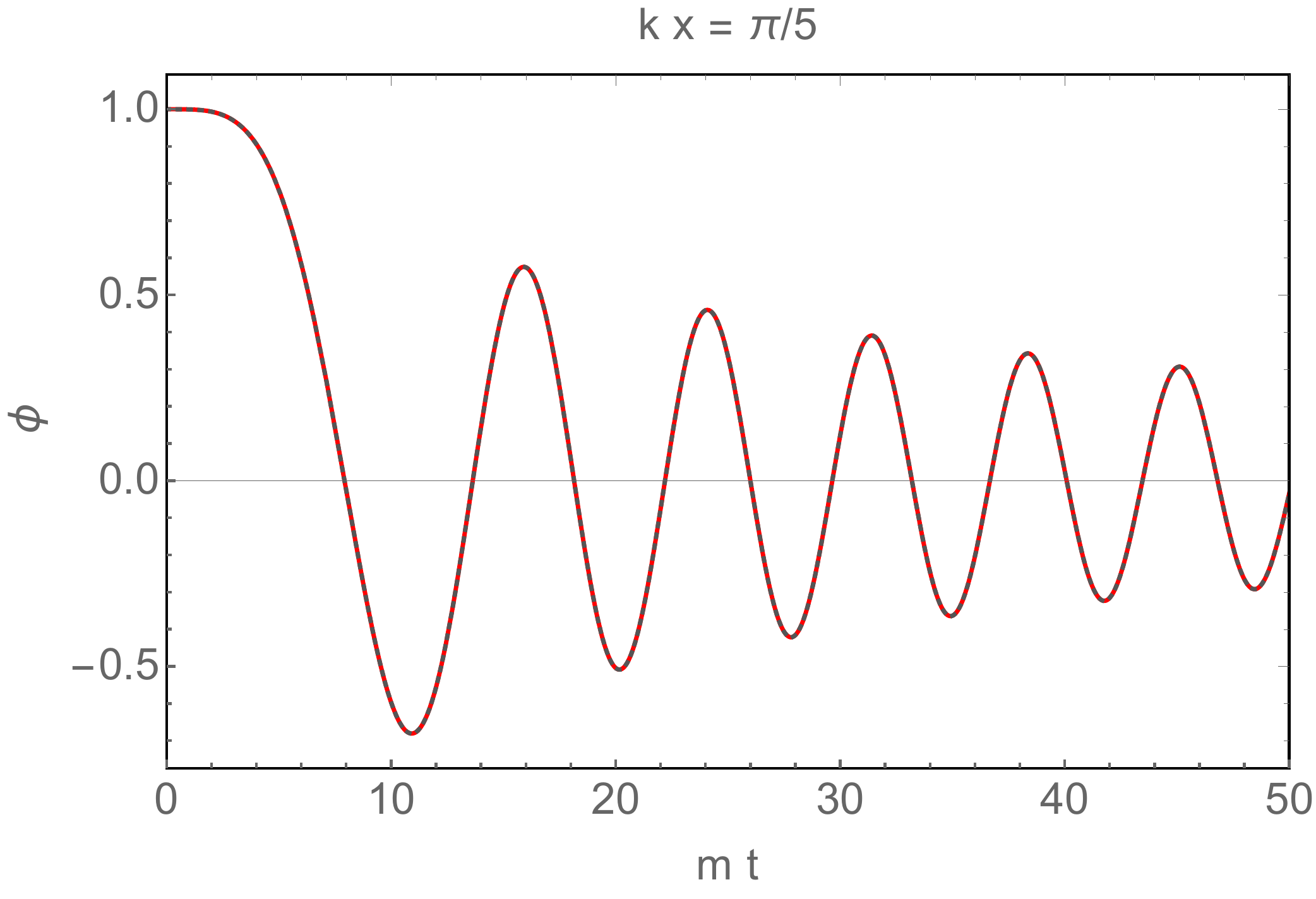}
	\includegraphics[width=0.32\textwidth]{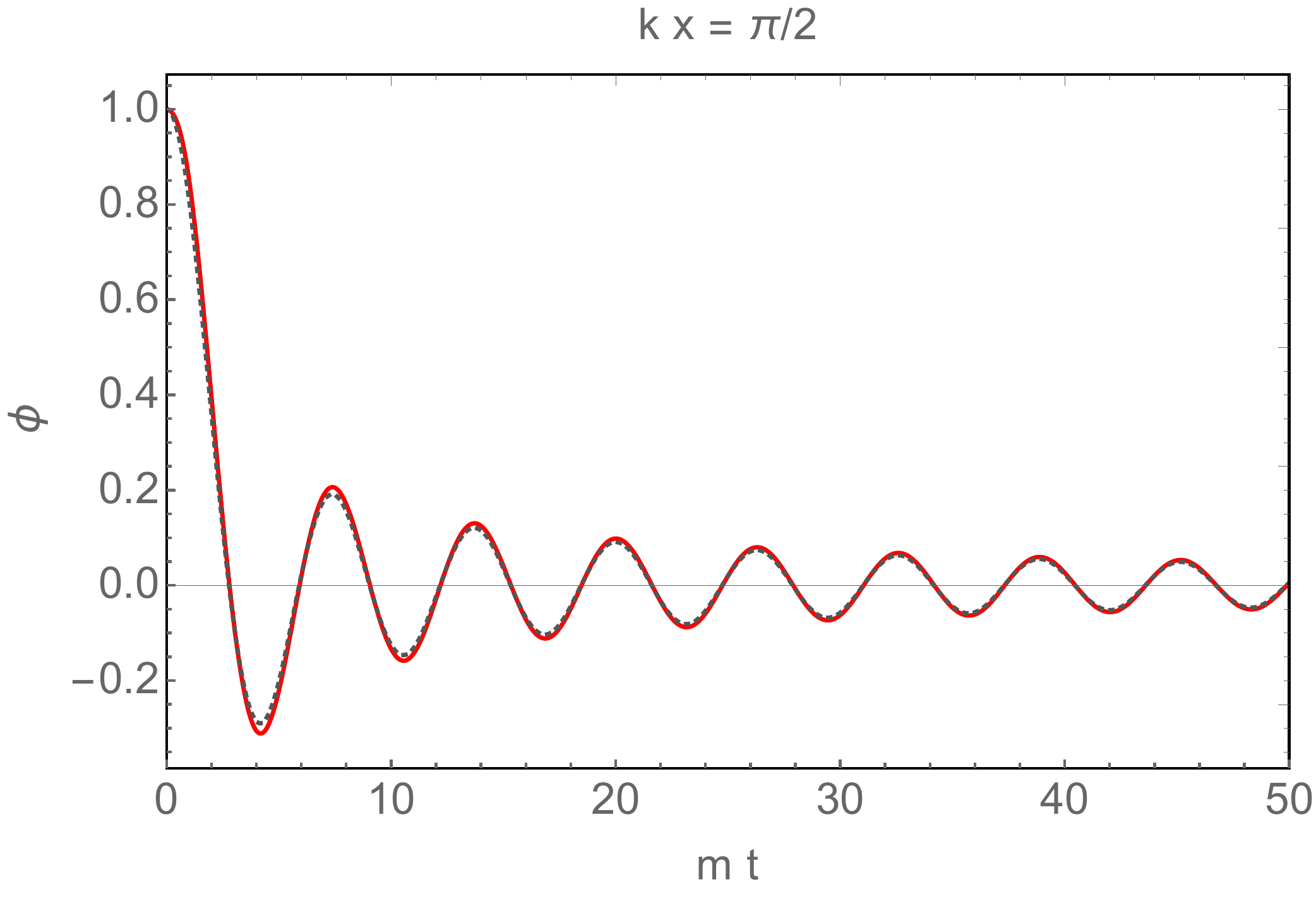}
	\caption{Comparing fixed $x=x_i$ slices of $\phi(t,x)$ in the small $k$ scenario (solid red curves) to the homogeneous case with the magnetic field given by $B \cos(kx_i)$ (dotted gray curves).}
	\label{fig:compare-homogeneous}
\end{figure}

\paragraph{$k \ll m < M$}
The scenario with $k$ smaller than the other mass scales also follows our expectations from the previous discussion.  The expectation is that different regions of space evolve approximately as if there is a homogeneous magnetic field, the value of which is determined by the local value of the magnetic field.  This behavior can be seen in Fig.~\ref{fig:compare-homogeneous} where we show the time evolution of $\phi(x,t)$ compared to a homogeneous case with $B = B(x)$ for three values of $x$.  There is an $\mathcal{O}(1)$ spatial change in the energy density in $\phi$ due to the change in magnetic fields with th periodicity given by $2k$.

\paragraph{$m < k \sim M$}
The spatial behavior of the intermediate case interpolates between the previous two scenarios. In Fig.~\ref{fig:rhoa} we show the evolution of the energy density in $\phi$ in all three cases compared to the homogeneous case. We see that results for small $k$ and intermediate $k$ are very similar to the result of the homogeneous case with an effective magnetic field $B_\text{eff} = B/\sqrt{2}$.  The energy density for the large $k$ case is smaller, as expected from the two state mixing approximation described in the previous section where the mass of the vector is identified with $k/a$ and thus the energy redshifts as $a^{-2}$ instead of $a^{-1}$ when $k/a > M$.

\end{document}